\documentclass[epj,final]{svjour}
\usepackage{graphicx}
\usepackage{dcolumn}
\usepackage{epsfig}
\usepackage{bm}
\usepackage[latin1]{inputenc}
\usepackage{amsmath}
\usepackage[T1]{fontenc}
\usepackage[cm]{aeguill}
\usepackage[english]{babel}
\usepackage{psfrag}
\usepackage{array}
\newcommand{\ff}[1]{\mbox{\boldmath$#1$}}

\newcommand{\comment}[1]{}
\usepackage{amsmath}
\usepackage{mathtools}
\usepackage{textcomp}

\begin{document}
\title{Numerical solutions of thin film equations for polymer flows}
\author{Thomas Salez\inst{1,}\thanks{e-mail: thomas.salez@espci.fr}\and Joshua D. McGraw\inst{2}\and Sara L. Cormier\inst{2}\and Oliver B\"{a}umchen\inst{2}\and Kari Dalnoki-Veress\inst{2}\and Elie Rapha\"{e}l\inst{1}}
\institute{Laboratoire de Physico-Chimie Th\'eorique, UMR CNRS Gulliver 7083, ESPCI, Paris, France
\and Department of Physics \& Astronomy and the Brockhouse Institute for Materials Research, McMaster University, Hamilton, Canada}
\date{\today}
\abstract{We report on the numerical implementation of thin film equations that describe the capillary-driven evolution of viscous films, in two-dimensional configurations. After recalling the general forms and features of these equations, we focus on two particular cases inspired by experiments: the leveling of a step at the free surface of a polymer film, and the leveling of a polymer droplet over an identical film. In each case, we first discuss the long-term self-similar regime reached by the numerical solution before comparing it to the experimental profile. The agreement between theory and experiment is excellent, thus providing a versatile probe for nanorheology of viscous liquids in thin film geometries.}
\maketitle

Even though hydrodynamics is a well established field, several limits and assumptions remain strongly debated. For instance, at which length scale and why does the theory break down? Can one generalize the concept of viscosity at small scales? What are the details of flow at the liquid/solid boundary? How does a droplet spread on a given wet or liquid substrate, and what is the role of the so-called precursor film? 

Polymeric liquids provide ideal systems with which to probe these fundamental questions. Molecular confinement effects can be investigated because of the large size of polymers and the ease with which one can reach system sizes that are comparable to that of the molecule. For instance, the enhancement of effective mobility has been predicted~\cite{Brochard2000}, observed~\cite{Bodiguel2006,Fakhraai2008}, and was related to a reduction of the effective entanglement density near surfaces~\cite{Si2005,Shin2007}. The relaxation to bulk mobility has also been observed for spin-coated polymer films~\cite{Reiter2001,Barbero2009,Raegen2010}. Moreover, larger molecular weights induce a larger slip at the substrate~\cite{deGennes1979,Baumchen2009}, which enables the study of slippery systems where the `no-slip' boundary condition must be modified \cite{Munch2005}. Another fundamental issue is the leveling of a polymer droplet onto an identical film. Whereas Tanner's regime of a droplet wetting a solid substrate through a thin precursor film is well understood~\cite{Tanner1979,deGennes1985,deGennes2003,Bonn2009}, the connection with the opposite regime where the droplet is thinner than the underlying film~\cite{Aradian2000,Pierce2009} should be explored further \cite{Cormier2012}. 

A common feature of these problems is that the liquids are viscous so that Stokes hydrodynamics may be applied \cite{Landau1987}. In particular, thin liquid films are well described by the lubrication theory through the thin film equations~\cite{Oron1997,Craster2009,Blossey2012}. However, apart from linearization~\cite{Bowen2006,Salez2012a}, these particular equations have not yet been solved analytically due to their high orders and non-linearities \cite{Bernis1990}. Mathematical advances can nevertheless be found in \cite{Myers1998,Kondic2003}. 

In the present communication, we focus on a numerical approach. In particular, we implement  the two-dimensional thin film equations for capillary-driven flows using schemes inspired by Bertozzi and Zhornitskaya~\cite{Bertozzi1998,Zhornitskaya2000}. After recalling the general forms and features of those thin film equations, we focus on two particular numerical situations that are compared to experimental systems: the leveling of a thin stepped polymer film~\cite{McGraw2011,McGraw2012} and the leveling of a polymer droplet on an identical polymer film~\cite{Cormier2012}. In both cases, we present the algorithm, the solution and its intermediate asymptotics that we compare to experimental data. Those intermediate regimes are of great interest for the non-linear thin film equations since they do not depend on the precise initial condition. Thus, they should be general intermediate solutions \cite{Barenblatt1996}.  

\section{General framework}
\label{genframe}
In this Section, we present the general thin film equations. After listing the common assumptions and boundary conditions, we address the particular cases where the film geometry is invariant along one horizontal or angular dimension.

As in~\cite{Cormier2012,McGraw2011,McGraw2012}, we consider polystyrene films above their glass transition temperature $T_{\textrm{g}}\sim100\ ^{\circ}\textrm{C}$. Extension to any other thin viscous liquid is straightforward using the relevant parameters. Here, we estimate the following typical parameters \cite{Wu1970,Rubinstein2003,Brandrup2005}: height $h_0\sim100$~nm, dynamical viscosity $\eta\sim1 \textrm{ MPa.s}$, molecular weight $M_\textrm{w}\sim100$~kg.mol$^{-1}$, surface tension $\gamma\sim30\ \textrm{mN.m}^{-1}$, density $\rho\sim1\ \textrm{g.cm}^{-3}$ and shear modulus $G\sim1$~MPa.
Let us evaluate the typical capillary velocity $v_{\textrm{c}}$, Reynolds number $\textrm{Re}$, capillary length $l_{\textrm{c}}$ and Maxwell viscoelastic time $\tau_\textrm{M}$:
\begin{subequations}
\begin{align}
v_{\textrm{c}}&=\frac{\gamma}{\eta}\sim2\ \mu\textrm{m.min}^{-1}\\
\textrm{Re}&=\frac{h_0\rho v_{\textrm{c}}}{\eta}\ll 1\\
l_{\textrm{c}}&=\sqrt{\frac{\gamma}{\rho g}}\sim 2\ \textrm{mm}\gg h_0\\
\tau_\textrm{M}&=\frac{\eta}{G}\sim1\ \textrm{s}\ .
\end{align}
\end{subequations}
With these parameters, and since we observe slow evolution of the liquid surface profile over several tens of minutes~\cite{Cormier2012,McGraw2011,McGraw2012}, we can make the following assumptions: we have an incompressible flow of a viscous Newtonian fluid where gravity~\cite{Huppert1982}, disjoining pressure~\cite{Seemann2001} and inertia are negligible. This flow is well described by the Stokes equation:
\begin{equation}
\label{stokes}
\ff{\nabla}P=\eta\ff{\Delta v}\ ,
\end{equation}
combined with the incompressibility condition:
\begin{equation}
\label{div}
\ff{\nabla\cdot v}=0\ ,
\end{equation}
where $P$ and $\ff{v}$ are the pressure and velocity fields within the liquid. In addition, we assume that the lubrication approximation is valid, that is: the profile slopes remain small in comparison to $1$. Finally, we assume $\gamma$ and $\eta$ to be homogeneous and constant.

As far as the vertical boundary conditions are concerned, we consider the case of no shear at the liquid-air interface:
\begin{equation}
\label{shear}
\partial_z \ff{v} |_{z=h}=\ff{0}\ ,
\end{equation}
where $z$ is the vertical coordinate, and we assume a no-slip boundary condition at the substrate:
\begin{equation}
\label{slip}
\ff{v_{\parallel}} |_{z=0}=\ff{0}\ ,
\end{equation}
where $v_{\parallel}$ is the projection of the velocity that is parallel to the liquid-substrate interface. Note that one could easily include slip at the substrate by imposing a nonzero $\ff{v_{\parallel}} |_{z=0}$ (see \textit{e. g.} \cite{Munch2005}).

Using the previous assumptions, we derive the thin film equations for two particular invariant geometries: transverse invariance (Case 1) and axisymmetry (Case 2).

In Case 1, we assume a spatial invariance of the problem in one horizontal direction $y$, which reduces the problem to two dimensions. The height of the profile is given by $h(x,t)$, where $x$ is the relevant horizontal direction and $t$ the time. The pressure is \textit{a priori} given by $P(x,z,t)$. According to the lubrication approximation, we can neglect the vertical velocities and write: $\ff{v}=v(x,z,t)\ \ff{e_x}$, where $\ff{e_x}$ is the horizontal basis vector. We then project and integrate Eq.~(\ref{stokes}), using Eq.~(\ref{div}), Eq.~(\ref{shear}) and Eq.~(\ref{slip}), and find:
\begin{equation}
\partial_z P=0\ ,
\end{equation}
the pressure $P(x,t)$ is thus invariant in the vertical direction, and:
\begin{equation}
\label{poiseuille}
v(x,z,t)=\frac{1}{2\eta}(z^2-2hz)\ \partial_x P\ ,
\end{equation} 
which corresponds to the familiar Poiseuille flow.
Volume conservation requires that:
\begin{equation}
\label{isov}
\partial_t h+\partial_x\int_0^{h}dz\ v=0\ .
\end{equation} 
Finally, because the pressure does not depend on $z$, we evaluate $P$ at the free surface through the Young-Laplace equation. Since the lubrication approximation implies small curvatures, the pressure satisfies:
\begin{equation}
\label{laplace}
P-P_0\approx-\gamma\partial_x^{\,2}h\ ,
\end{equation} 
where $P_0$ is the atmospheric pressure.
Thus, combining Eq.~(\ref{poiseuille}), Eq.~(\ref{isov}) and Eq.~(\ref{laplace}), we get:
\begin{equation}
\label{fulltfe}
\partial_th+\frac{\gamma}{3\eta}\partial_x(h^3\partial_x^{\,3}h)=0\ ,
\end{equation} 
which is the general capillary-driven thin film equation. 

In Case 2, we assume an invariance of the problem by rotation about the vertical axis. Therefore, the height of the profile is given by $h(r,t)$, where $r$ is the radial coordinate. The pressure is \textit{a priori} given by $P(r,z,t)$. According to the lubrication approximation, we can neglect the vertical velocities and write: $\ff{v}=v(r,z,t)\ \ff{e_r}$, where $\ff{e_r}$ is the radial basis vector in cylindrical coordinates. By proceeding similarly to Case 1, and including the two principal curvatures, we obtain:
\begin{equation}
\label{ctfe}
\partial_t h+\frac{\gamma}{3\eta}\frac{1}{r}\partial_r\left[rh^3\left(\partial_r^{\,3}h+\frac{1}{r}\partial_r^{\,2}h-\frac{1}{r^2}\partial_rh\right)\right]=0\ .
\end{equation} 
It is worth stressing that although this axisymmetric geometry has been studied in the past through a similar equation \cite{Stillwagon1988,Stillwagon1990}, Eq.~(\ref{ctfe}) is more general since it includes the two principal curvatures with no far field approximation.

\section{Leveling of a stepped polymer film}
\label{sectfe}
\begin{figure}
\begin{center}
\includegraphics[width=8.7cm]{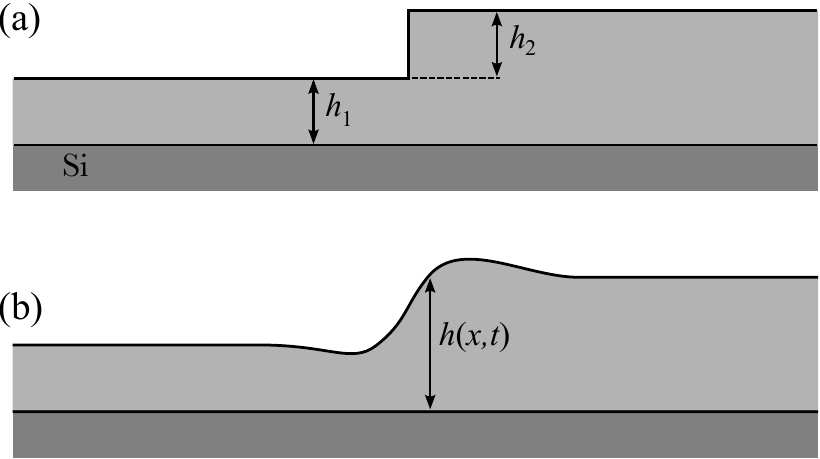}
\caption{\textit{Schematic of the experiment. The ensemble is placed on a solid silicon (Si) substrate. (a) As-prepared stepped film in the glassy state~\cite{McGraw2011,McGraw2012}. The heights are of the order of $h_1\sim h_2\sim100$~nm. (b) Above the glass transition temperature, the profile $h(x,t)$ levels with time due to capillary-driven viscous flow.}}
\label{step}
\end{center}
\end{figure}

In this Section, we focus on the experimental situation described in~\cite{McGraw2011,McGraw2012}: a polystyrene \textit{stepped film} with initial heights $h_1$ and $h_2$ (see Fig.~\ref{step}), which levels above the glass transition temperature due to the capillary-driven viscous flow. In the following, we present the dimensionless mathematical model before describing the algorithm. We then characterize the numerical solution. In particular, we study the long-term self-similarity of the evolution and the sensitivity of this regime to initial conditions. Finally, we compare the results to experimental data. 

\subsection{Mathematical model}
Since this problem is spatially invariant along one horizontal direction, we refer to Case 1 introduced in Section~\ref{genframe}. We are interested in the total height of the free surface $h(x,t)$ (see Fig.~\ref{step}). Let us introduce the natural dimensionless variables:
\begin{subequations}
\label{natl}
\begin{align}
\label{natlength}
H&=\frac{h}{h_0}\\
\label{natlength2}
X&=\frac{x}{h_0}\\
\label{natlength3}
T&=\frac{\gamma t}{3\eta h_0}\ ,
\end{align}
\end{subequations}
and the dimensionless parameters:
\begin{subequations}
\label{natl2}
\begin{align}
H_1&=\frac{h_1}{h_0}\\
H_2&=\frac{h_2}{h_0}\ ,
\end{align}
\end{subequations}
where $h_0$ is a chosen length scale of reference. Then, we non-dimensionalize Eq.~(\ref{fulltfe}) to obtain:
\begin{equation}
\label{adtfe}
\partial_T H+\partial_X\left(H^3\partial_X^{\,3}H\right)=0\ .
\end{equation}
In addition, we consider the following dimensionless initial condition:
\begin{align}
H(X,0) =
\begin{dcases}
H_1, \quad &X<0\\
H_1+H_2, \quad &X>0\\
H_1+\frac{H_2}{2}, \quad &X=0 \ ,
\end{dcases}
\label{ci}
\end{align}
according to Fig.~\ref{step}, Eq.~(\ref{natl}) and Eq.~(\ref{natl2}).

\subsection{Algorithm}
\label{algo}
The numerical procedure we use is a finite difference method developed in \cite{Bertozzi1998,Zhornitskaya2000}. It ensures capillary energy and entropy dissipation as required from \cite{Bernis1990}. In addition to volume conservation, it has been shown in~\cite{Zhornitskaya2000} that this method ensures positivity of the height profile $H(X,T)$. In the following, we describe the discretization as well as the initial and boundary conditions of our algorithm, before presenting the integration routine.

We discretize space through the definition:
\begin{equation}
\mathcal{H}_i=H\left[\left(i-\frac{M+1}{2}\right)\Delta X,T\right]\ ,
\end{equation}
for $i\in[1,M]$, where $\Delta X$ is the dimensionless spatial increment. We introduce the growth rates:
\begin{subequations}
\begin{align}
\mathcal{H}_{X,i}&=\frac{\mathcal{H}_{i+1}-\mathcal{H}_i}{\Delta X}\\
\mathcal{H}_{\bar{X},i}&=\mathcal{H}_{X,i-1}\\
\mathcal{H}_{\bar{X}X,i}&=\frac{\mathcal{H}_{\bar{X},i+1}-\mathcal{H}_{\bar{X},i}}{\Delta X}\\
\mathcal{H}_{\bar{X}X\bar{X},i}&=\frac{\mathcal{H}_{\bar{X}X,i}-\mathcal{H}_{\bar{X}X,i-1}}{\Delta X}\ .
\end{align}
\end{subequations}
Then, we consider a continuous-time discrete-space approximation of Eq.~(\ref{adtfe}):
\begin{equation}
\label{dtfe}
\frac{d\mathcal{H}_i}{dT} =\frac{\mathcal{A}_i\mathcal{H}_{\bar{X}X\bar{X},i}-\mathcal{A}_{i+1}\mathcal{H}_{\bar{X}X\bar{X},i+1}}{\Delta X}\ ,
\end{equation}
with $i\in[1,M-1]$, and where we define the auxiliary function $\mathcal{A}_i$ as:
\begin{equation}
\label{aux}
\mathcal{A}_i=2\frac{\mathcal{H}_{i-1}^{\,2}\mathcal{H}_i^{\,2}}{\mathcal{H}_{i-1}+\mathcal{H}_i}\ ,
\end{equation}
for $i\in[1,M]$. Note that for small slopes and small spatial increments: $\mathcal{A}_i\approx \mathcal{H}_i^{\,3}$, as expected from standard finite difference method applied to Eq.~(\ref{adtfe}).
\begin{table}[b!]
\begin{center}
\begin{tabular}{|l|l|}
\hline
 & \\
Parameter & Value  \\
\hline
&  \\
$\Delta T$ & $1.0\times10^{-6}$\\
$\Delta X$ & $1.0\times 10^{-1}$\\
$N$ & $10000000$\\
$M$ & $501$\\
$H_1$ & $1.0$\\
$H_2$ & $1.0$\\
&  \\
\hline
\end{tabular}
\caption{Optimized parameters for the numerical resolution of Eq.~(\ref{dtfe}).
\label{tabTFE}}
\end{center}
\end{table}

For the initial condition ($T=0$), we use the discretized version of Eq.~(\ref{ci}):
\begin{align}
\mathcal{H}_i =
\begin{dcases}
H_1, \quad &1\leq i<\frac{M+1}{2}\\
H_1+H_2, \quad &\frac{M+1}{2}<i\leq M\\
H_1+\frac{H_2}{2}, \quad &i=\frac{M+1}{2}\ .
\end{dcases}
\label{numci}
\end{align}

For integration, since we have a finite numerical window, we choose the following horizontal boundary conditions at $T>0$:
\begin{subequations}
\begin{align}
\label{aux1}
\mathcal{A}_1&=\mathcal{H}_1^{\,3}\\
\mathcal{H}_{\bar{X},1}&=0\\
\mathcal{H}_{\bar{X}X,M}&=0\\
\mathcal{H}_{\bar{X}X\bar{X},1}&=0\\
\frac{d\mathcal{H}_M}{dT}&=\frac{\mathcal{A}_M\mathcal{H}_{\bar{X}X\bar{X},M}}{\Delta X}\ .
\end{align}\end{subequations} 
Note that the spatial window size $M\Delta X$ must be chosen large enough in order to have relevant horizontal boundary conditions.

Finally, we solve Eq.~(\ref{dtfe}) for any $T>0$ using a fourth order Runge-Kutta routine \cite{Press1996}. Time is  discretized through $T=j\Delta T$, where $j\in[1,N]$, and where $\Delta T$ is the dimensionless temporal increment. Note that, according to~\cite{Stillwagon1988}, we should have at least $\Delta T\sim\Delta X^4$ due to the general orders of Eq.~(\ref{adtfe}). The typical numerical parameters after optimization are summarized in Table~\ref{tabTFE}.

\subsection{Results}
In this part, we present the numerical solution of Eq.~(\ref{adtfe}) using the algorithm presented above. We characterize the long-term self-similarity of the evolution as well as the robustness of this regime with respect to variations of the initial condition. 

\begin{figure}
\begin{center}
\includegraphics[width=8.8cm]{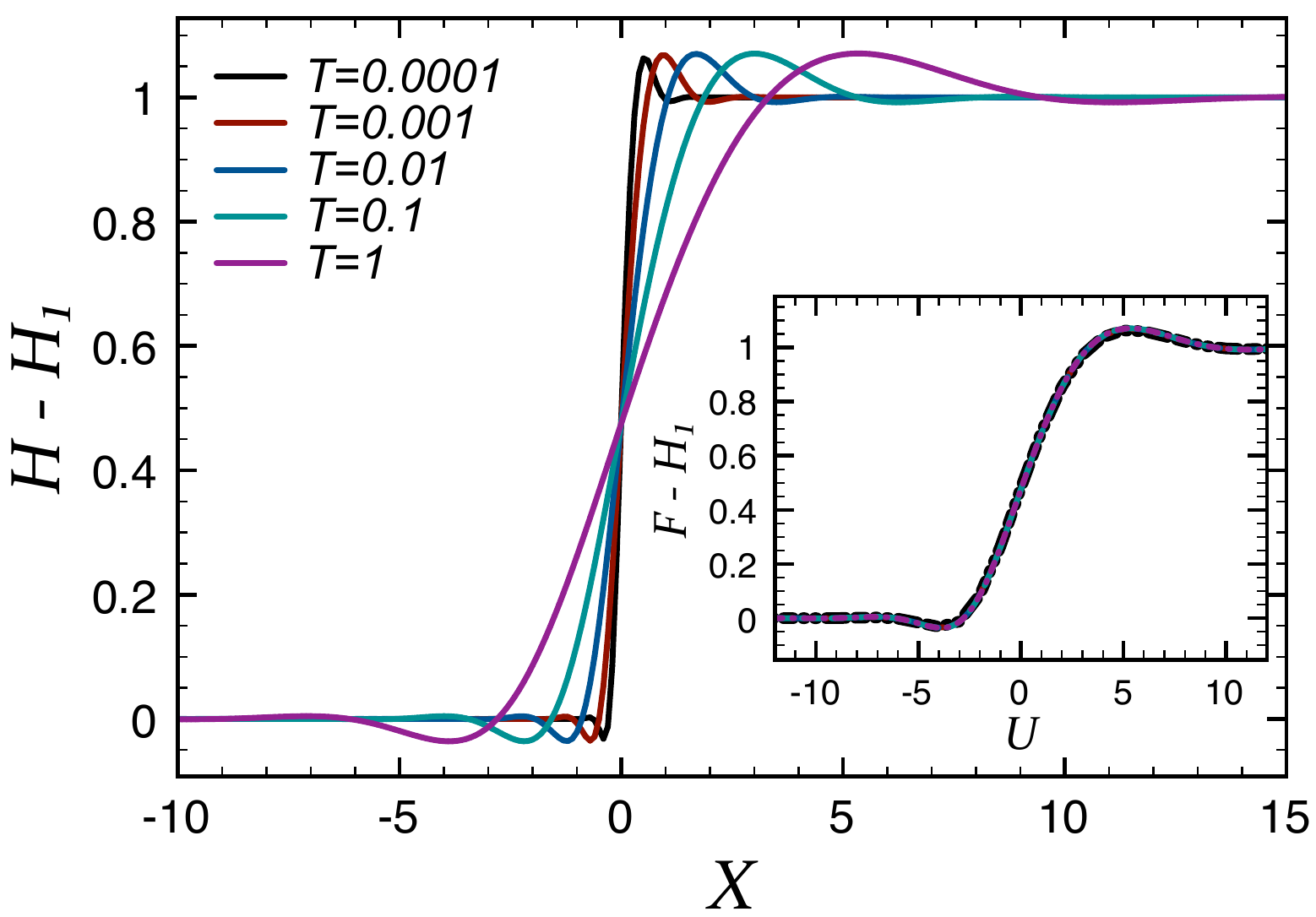}
\caption{\textit{Numerical solution of Eq.~(\ref{adtfe}) at different dimensionless  times. The initial step is chosen such that $H_1=H_2=1$. The inset shows a self-similar replotting of the data according to Eq.~(\ref{eqss}).}}
\label{ssplot1}
\end{center}
\end{figure}

As expected, the initial step levels due to the gradients in capillary pressure. The numerical solution $H(X,T)-H_1$ is plotted at different dimensionless times in Fig.~\ref{ssplot1}. 
The initial step has been chosen such that $H_1=H_2=1$. In addition, we observe spatially damped oscillations of the free surface. 

Guided by the symmetry of Eq.~(\ref{adtfe}) and Eq.~(\ref{ci}), we look for self-similar solutions of the first kind \cite{Barenblatt1996,Aradian2001} defined by:
\begin{subequations}
\label{eqss}
\begin{align}
F(U)&=H(X,T)\\
U&=\frac{X}{T^{1/4}}\ .
\end{align}
\end{subequations}
We thus replot the evolution of Fig.~\ref{ssplot1} in this new variable. The result is shown in inset of the same figure. All the curves collapse onto a single curve, demonstrating the self-similarity of the long-term evolution. However, this intermediate asymptotics appears to be reached only at large times, as in \cite{Christov2012}. During the short-term evolution, the oscillations are observed to grow with time before saturating.
\begin{figure}
\begin{center}
\includegraphics[width=8.8cm]{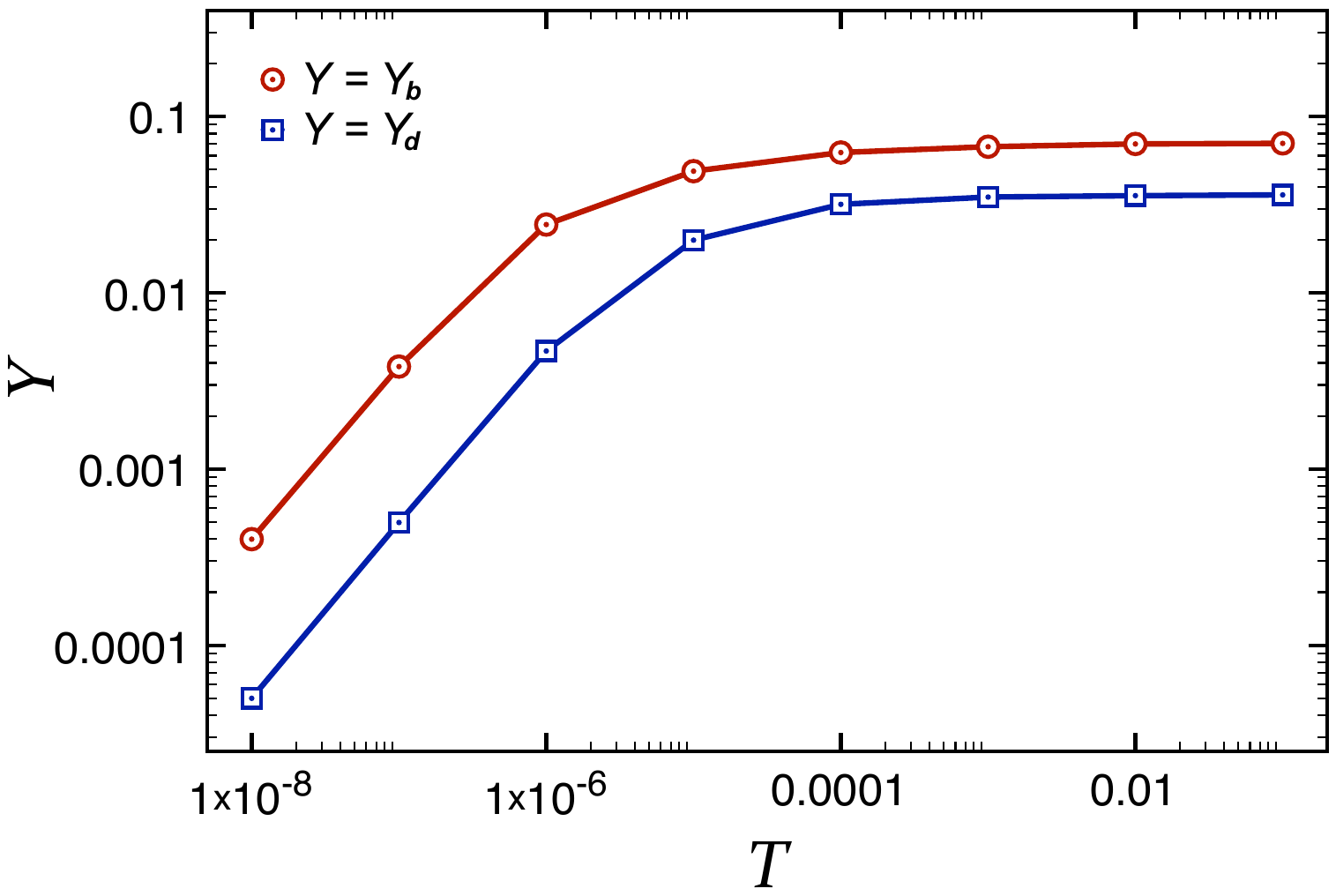}
\caption{\textit{Evolution of the bump and the dip of the numerical solution defined in Eq.~(\ref{bdef}), for an initial profile with $H_1=H_2=1$.}}
\label{bd}
\end{center}
\end{figure}
In order to quantify this statement, we study the \textit{bump} and the \textit{dip} of the leveling profile:
\begin{subequations}
\label{bdef}
\begin{align}
Y_\textrm{b}(T)&= \left|\max_X[H(X,T)]-\max_X[H(X,0)] \right|\\
Y_\textrm{d}(T)&=\left|\min_X[H(X,T)]-\min_X[H(X,0)]\right|\ .
\end{align}
\end{subequations}
Defined in this way, the bump and the dip are the extremal heights above and below the as-prepared step profile, respectively. The temporal evolutions of $Y_\textrm{b}$ and $Y_\textrm{d}$ for a film with initial condition $H_1=H_2=1$ are shown in Fig.~\ref{bd}. They saturate after $T\sim0.0001$, indicating that the self-similar regime is reached.

\begin{figure}
\begin{center}
\includegraphics[width=8.4cm]{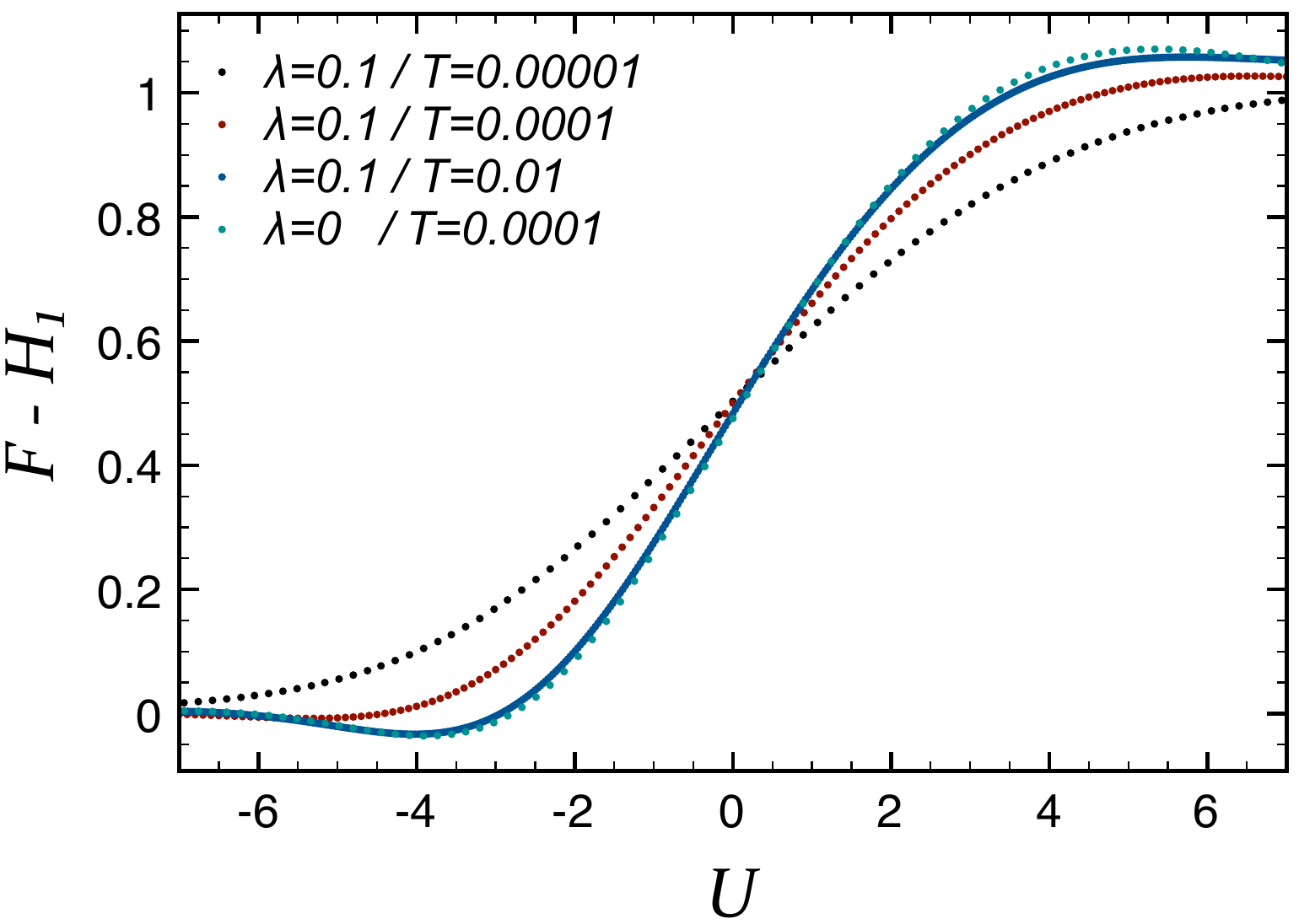}
\caption{\textit{Numerical solution for a smooth initial condition given by Eq.~(\ref{dirac}) with $H_1=H_2=1$, $\lambda=0.1$ and $\Delta X=0.01$. The profile is plotted at several dimensionless times, in self-similar variables according to Eq.~(\ref{eqss}). For comparison, the solution from a pure Heaviside initial condition ($\lambda=0$) has been plotted as well.}}
\label{testin}
\end{center}
\end{figure}

In order to understand the stability of the self-similar regime described in the previous Section, we turned the sharp initial profile to a smoother Fermi-Dirac function:
\begin{equation}
\label{dirac}
H_\lambda(X,0)=H_1+\frac{H_2}{1+\exp\left(-\frac{X}{\lambda}\right)}\ ,
\end{equation}
where $\lambda$ is the width of the transition region. Note that it is necessary to adapt the spatial increment so that the condition $\lambda\gg \Delta X$ is fulfilled. The results for $\lambda=0.1$ and $\lambda=0$ are compared in Fig.~\ref{testin}.
Both initial conditions converge to the same self-similar regime of Eq.~(\ref{eqss}). The smaller $\lambda$ is, the faster the self-similar regime is reached as expected according to the driving Laplace's pressure of Eq.~(\ref{laplace}). The asymptotic regime is thus attractive and robust with respect to a variation of the initial profile at constant boundary limits. This result is important for comparison to experiments since it validates \textit{a posteriori} the contradictory use of a sharp stepped initial condition in Eq.~(\ref{ci}) within the lubrication approximation of small slopes underlying Eq.~(\ref{adtfe}). Moreover, this tells us that experimentally fabricated samples with their inherent imperfections should still approach the theoretical self-similar regime, as discussed below.

\subsection{Comparison with experiments}
\begin{figure}[t!]
\begin{center}
\includegraphics[width=8.3cm]{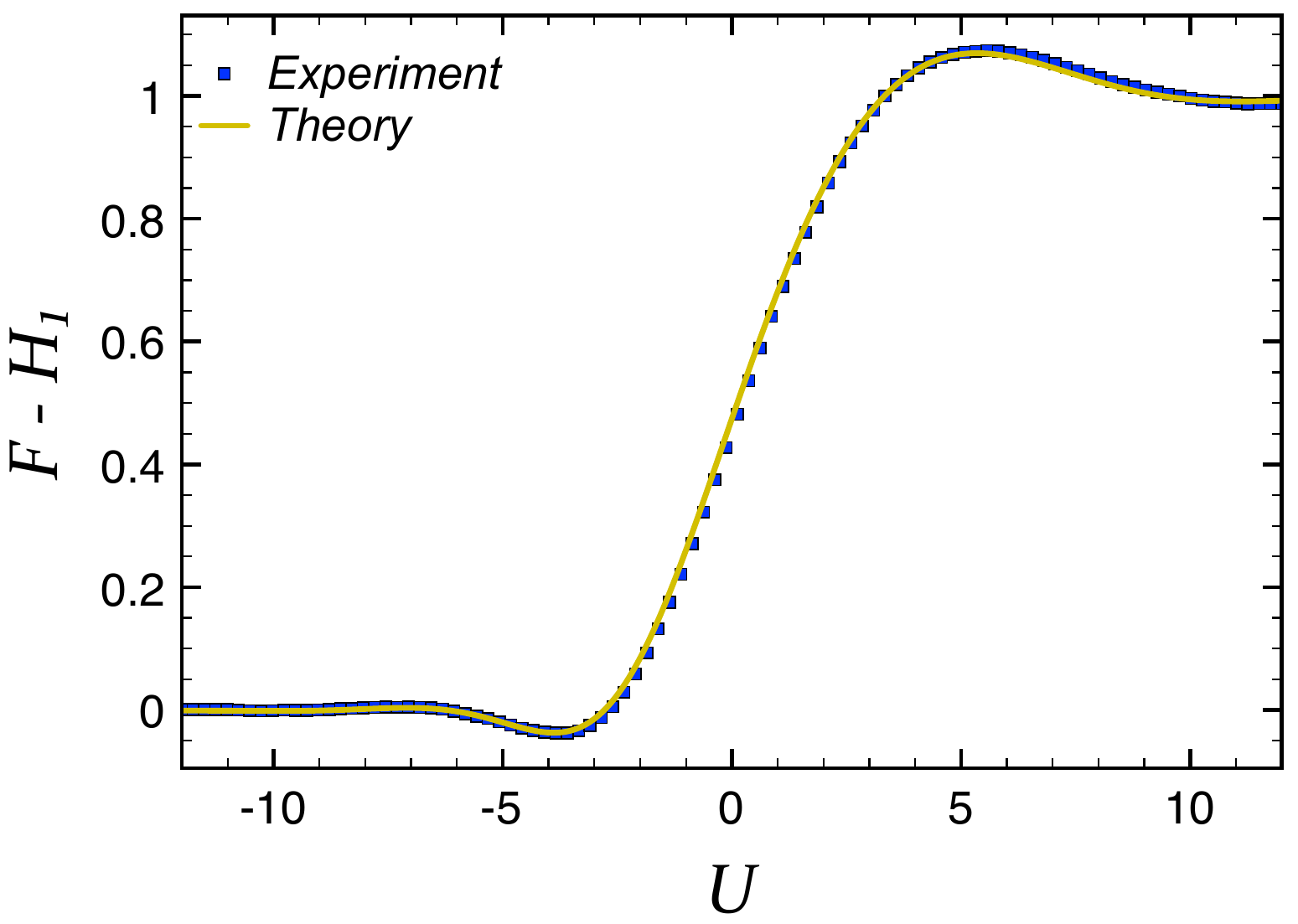}
\caption{\textit{Fit of the numerical solution of Eq.~(\ref{adtfe}) to experimental data (atomic force microscopy), in self-similar variables according to Eq.~(\ref{eqss}). The system is a  polystyrene stepped film, with $M_\textrm{w}=118$~kg.mol$^{-1}$ and $h_1=h_2=89$~nm, leveling on a silicon substrate~\cite{McGraw2011,McGraw2012}. The temperature of the sample is maintained at $140^\circ\textrm{C}$.}}
\label{resbilayer}
\end{center}
\end{figure}

In order to demonstrate the interest in such a numerical solution, we now compare it with experiments. Details can be found in~\cite{McGraw2011,McGraw2012}. Note that the self-similarity of the experimental profiles has been demonstrated in~\cite{McGraw2012}. Figure~\ref{resbilayer} shows a comparison between the numerical solution of Eq.~(\ref{adtfe}) and an experimental profile. The sample is a polystyrene stepped film ($h_1=h_2=89$~nm) with $M_\textrm{w}=118$~kg.mol$^{-1}$, annealed at $140^\circ\textrm{C}$. 
As we can see, the agreement is excellent, with residuals being less than one percent of the data range. The vertical scaling parameter is determined by $h_0$ according to Eq.~(\ref{natl}) and Eq.~(\ref{natl2}). Thus, if the capillary velocity $\gamma/\eta$ is unknown at the considered temperature, the single fitting parameter is the horizontal stretch defined by:
\begin{subequations}
\label{ufit}
\begin{align}
U&=\frac{X}{T^{1/4}}\\
&=\left(\frac{3\eta}{\gamma h_0^{\,3}}\right)^{1/4}\frac{x}{t^{1/4}}\ ,
\end{align}
\end{subequations}
according to Eq.~(\ref{natl}). For the data shown in Fig.~\ref{resbilayer}, we find $\gamma/\eta= 1.5 \pm 0.3\ \textrm{\textmu m.min}^{-1}$ at $140^\circ\textrm{C}$ \cite{McGraw2012}, which compares well with the tabulated values~\cite{Wu1970,Bach2003}, through the WLF model~\cite{Williams1955}.
This agreement validates the numerical results and the methodology itself. Moreover, the technique appears to provide an accurate measurement of the capillary velocity of the liquid. 

\section{Leveling of a polymer droplet on an identical film}
\label{secctfe}
\begin{figure}
\begin{center}
\includegraphics[width=8.8cm]{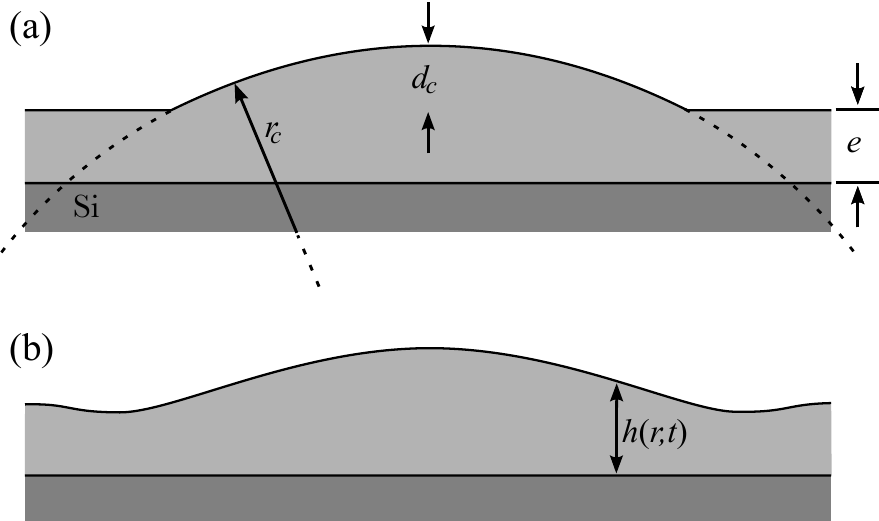}
\caption{\textit{Schematic of the experimental configuration in \cite{Cormier2012}. The ensemble is placed on a solid silicon (Si) substrate. (a) A thin polystyrene droplet, of initial height $d_\textrm{c}$ and radius of curvature $r_\textrm{c}$, is placed onto an identical film of thickness $e$, in the glassy state. (b) After heating above the glass transition temperature, one observes the capillary leveling of the profile $h(r,t)$.}}
\label{dropsketch}
\end{center}
\end{figure}

In this Section, we address a second problem inspired by experiments: the leveling of a thin liquid polymer droplet on an identical film~\cite{Cormier2012}. We assume that the initial shape of the droplet is a spherical cap of height $d_\textrm{c}$ and radius of curvature $r_\textrm{c}$, and that the underlying film of thickness $e$ is infinitely wide (see Fig.~\ref{dropsketch}). 
Whereas Tanner's regime of a droplet wetting a solid substrate is well understood by invoking the presence of an infinitesimal precursor film~\cite{Tanner1979,deGennes1985,deGennes2003,Bonn2009}, the connection towards the opposite regime~\cite{Aradian2000,Pierce2009,Cormier2012}, where the droplet is thinner than the film, is of interest. In the following, we establish the dimensionless mathematical model before presenting our numerical algorithm. Then, we characterize the resulting numerical solution. In particular, we study the long-term self-similarity of this evolution. Finally, we compare the results to experimental data. 

\subsection{Mathematical model}
Since this problem is axisymmetric, we refer to Case 2 introduced in Section~\ref{genframe}. We are interested in the total height of the free surface $h(r,t)$ (see Fig.~\ref{dropsketch}). Let us introduce the natural dimensionless variables:
\begin{subequations}
\label{natlctfe}
\begin{align}
R&=\frac{r}{h_0}\\
H&=\frac{h}{h_0}\\
T&=\frac{\gamma t}{3\eta h_0} \ ,
\end{align}
\end{subequations}
and the dimensionless parameters:
\begin{subequations}
\label{natlctfe2}
\begin{align}
D_\textrm{c}&=\frac{d_\textrm{c}}{h_0}\\
R_\textrm{c}&=\frac{r_\textrm{c}}{h_0}\\
E&=\frac{e}{h_0}\ ,
\end{align}
\end{subequations}
where $h_0$ is a chosen length scale of reference. Then, we non-dimensionalize Eq.~(\ref{ctfe}) to obtain:
\begin{equation}
\label{adctfe}
\partial_T H+\frac{1}{R}\partial_R\left[RH^3\left(\partial_R^{\,3}H+\frac{1}{R}\partial_R^{\,2}H-\frac{1}{R^2}\partial_RH\right)\right]=0\ .
\end{equation}
In addition, we consider the following dimensionless initial condition:
\begin{align}
\label{cictfe}
H(R,0)&=
\begin{dcases}
E+D_\textrm{c}-R_\textrm{c}+\sqrt{R_\textrm{c}^{\,2}-R^2}, \quad &R\leq A\\
E, \quad &R>A\ ,
\end{dcases}
\end{align}
according to Fig.~\ref{dropsketch}, Eq.~(\ref{natlctfe}) and Eq.~(\ref{natlctfe2}). We introduced the distance $A$ from the center to the initial point where the flat film intersects the droplet profile:
\begin{align}
A&=\sqrt{2R_\textrm{c}D_\textrm{c}-D_\textrm{c}^{\,2}}\ .
\end{align}

\subsection{Algorithm}
We use a similar algorithm as the one introduced in Part~\ref{algo}. In the following, we describe the discretization as well as the initial and boundary conditions, before presenting the integration routine.
\begin{table}[b!]
\begin{center}
\begin{tabular}{|l|l|}
\hline
 & \\
Parameter & Value  \\
\hline
&  \\
$\Delta T$ & $1.0\times10^{-10}$\\
$\Delta R$ & $1.0\times 10^{-2}$\\
$N$ & $100000000$\\
$M$ & $300$\\
$D_\textrm{c}$ & $0.1$\\
$R_\textrm{c}$ & $2.0$\\
$E$ & $1.0$\\
&  \\
\hline
\end{tabular}
\caption{Optimized parameters for the numerical resolution of Eq.~(\ref{dctfe}).
\label{tabCTFE}}
\end{center}
\end{table}

For spatial discretization, we define:
\begin{equation}
\mathcal{H}_i=H\left[\left(i-\frac{M+1}{2}\right)\Delta R,T\right]\ ,
\end{equation} 
for $i\in[1,M]$, where $\Delta R$ is the dimensionless spatial increment. We introduce the growth rates:
\begin{subequations}
\begin{align}
\mathcal{H}_{R,i}&=\frac{\mathcal{H}_{i+1}-\mathcal{H}_i}{\Delta R}\\
\mathcal{H}_{\bar{R},i}&=\mathcal{H}_{R,i-1}\\
\mathcal{H}_{\bar{R}R,i}&=\frac{\mathcal{H}_{\bar{R},i+1}-\mathcal{H}_{\bar{R},i}}{\Delta R}\\
\mathcal{H}_{\bar{R}R\bar{R},i}&=\frac{\mathcal{H}_{\bar{R}R,i}-\mathcal{H}_{\bar{R}R,i-1}}{\Delta R}\ .
\end{align}
\end{subequations}  
Then, using Eq.~(\ref{aux}), we consider a continuous-time discrete-space approximation of Eq.~(\ref{adctfe}):
\begin{eqnarray}
\label{dctfe}
\frac{d\mathcal{H}_i}{dT} &=&\frac{\mathcal{A}_i\mathcal{H}_{\bar{R}R\bar{R},i}-\mathcal{A}_{i+1}\mathcal{H}_{\bar{R}R\bar{R},i+1}}{\Delta R}\nonumber\\
&+&\frac{\mathcal{A}_i\mathcal{H}_{\bar{R}R,i}-\mathcal{A}_{i+1}\mathcal{H}_{\bar{R}R,i+1}}{(i-3)\Delta R^2}\nonumber\\
&+&\frac{\mathcal{A}_{i+1}\mathcal{H}_{\bar{R},i+1}-\mathcal{A}_i\mathcal{H}_{\bar{R},i}}{(i-3)^2\Delta R^3}\nonumber\\
&-&\frac{\mathcal{A}_i\mathcal{H}_{\bar{R}R\bar{R},i}}{(i-3)\Delta R}-\frac{\mathcal{A}_i\mathcal{H}_{\bar{R},i}}{(i-3)^3\Delta R^3}\ ,
\end{eqnarray}
with $i\in[4,M-1]$.

At $T=0$, we use the discretized version of Eq.~(\ref{cictfe}) as an initial condition:
\begin{align}
\label{sphereci}
\mathcal{H}_{i}=
\begin{dcases}
E+D_\textrm{c}-R_\textrm{c}+\sqrt{R_\textrm{c}^{\,2}-(i-3)^2\Delta R^2}, \quad & i		\leq k\\
E, \quad & i>k\ ,
\end{dcases}
\end{align}
with:
\begin{align}
k&=3+\left\lfloor\frac{\sqrt{2R_\textrm{c}D_\textrm{c}-D_\textrm{c}^{\,2}}}{\Delta R}\right\rfloor\ ,
\end{align}
where we introduced the floor notation $\lfloor \,\rfloor$.

For integration, since we have a finite numerical window, we choose the following horizontal boundary conditions at $T>0$:
\begin{subequations}
\begin{align}
\mathcal{H}_1&=\mathcal{H}_5\\
\mathcal{H}_2&=\mathcal{H}_4\\
\mathcal{H}_{\bar{R}R,M}&=0\\
\frac{d\mathcal{H}_M}{dT} &=0\\
\frac{d\mathcal{H}_3}{dT} &=\frac{8}{3}\ \frac{\mathcal{A}_3\mathcal{H}_{\bar{R}R\bar{R},3}-\mathcal{A}_{4}\mathcal{H}_{\bar{R}R\bar{R},4}}{\Delta R}\ .
\end{align}
\end{subequations} 
The last equation corresponds to the limit of Eq.~(\ref{adctfe}) when $R\rightarrow0$. The value $i'=0$ corresponds to $R=0$, since we need two additional points to define the central growth rates. Again, we note that the spatial window size $M\Delta R$ must be chosen large enough in order to have relevant horizontal boundary conditions. 

Finally, we solve Eq.~(\ref{dctfe}) for any $T>0$ using a fourth order Runge-Kutta routine \cite{Press1996}. We discretize time through $T=j\Delta T$, where $j\in[1,N]$ and where $\Delta T$ is the dimensionless temporal increment. Note that, according to~\cite{Stillwagon1988}, we should have at least $\Delta T\sim\Delta R^4$ due to the general orders of Eq.~(\ref{adctfe}). The typical numerical parameters after optimization are summarized in Table~\ref{tabCTFE}.
\begin{figure}[t!]
\begin{center}
\includegraphics[width=8.7cm]{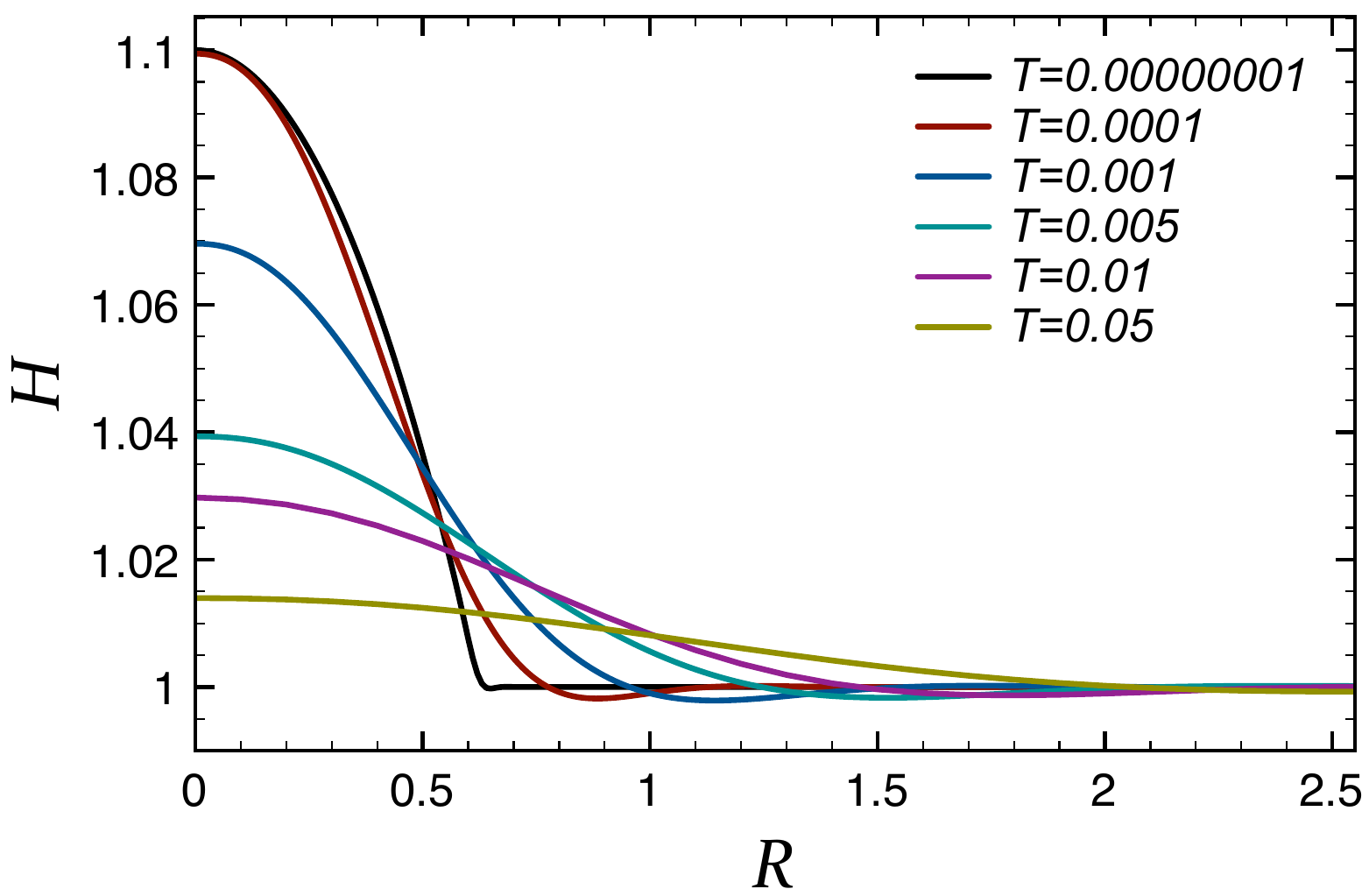}
\caption{\textit{Numerical solution of Eq.~(\ref{adctfe}), after several dimensionless times. The initial condition is a spherical cap of dimensionless height $D_\textrm{c}=0.1$ and radius of curvature $R_\textrm{c}=2$, on a film with height $E=1$.}}
\label{dropfall}
\end{center}
\end{figure} 

\subsection{Results}
In this Part, we present the numerical solution of Eq.~(\ref{adctfe}) using the algorithm introduced above. Then, we characterize the long-term self-similarity of the evolution and the convergence to this regime.

As expected the initial drop spreads due to the large gradients in capillary pressure. The numerical solution $H(R,T)$ is plotted in Fig.~\ref{dropfall} at different dimensionless times. 
Here, the initial condition is a spherical cap of dimensionless height $D_\textrm{c}=0.1$ and radius of curvature $R_\textrm{c}=2$, on a film with height $E=1$. Note the presence of oscillations of the free surface, and especially a dip, as in Fig.~\ref{ssplot1}. 

\begin{figure}
\begin{center}
\includegraphics[width=8.0cm]{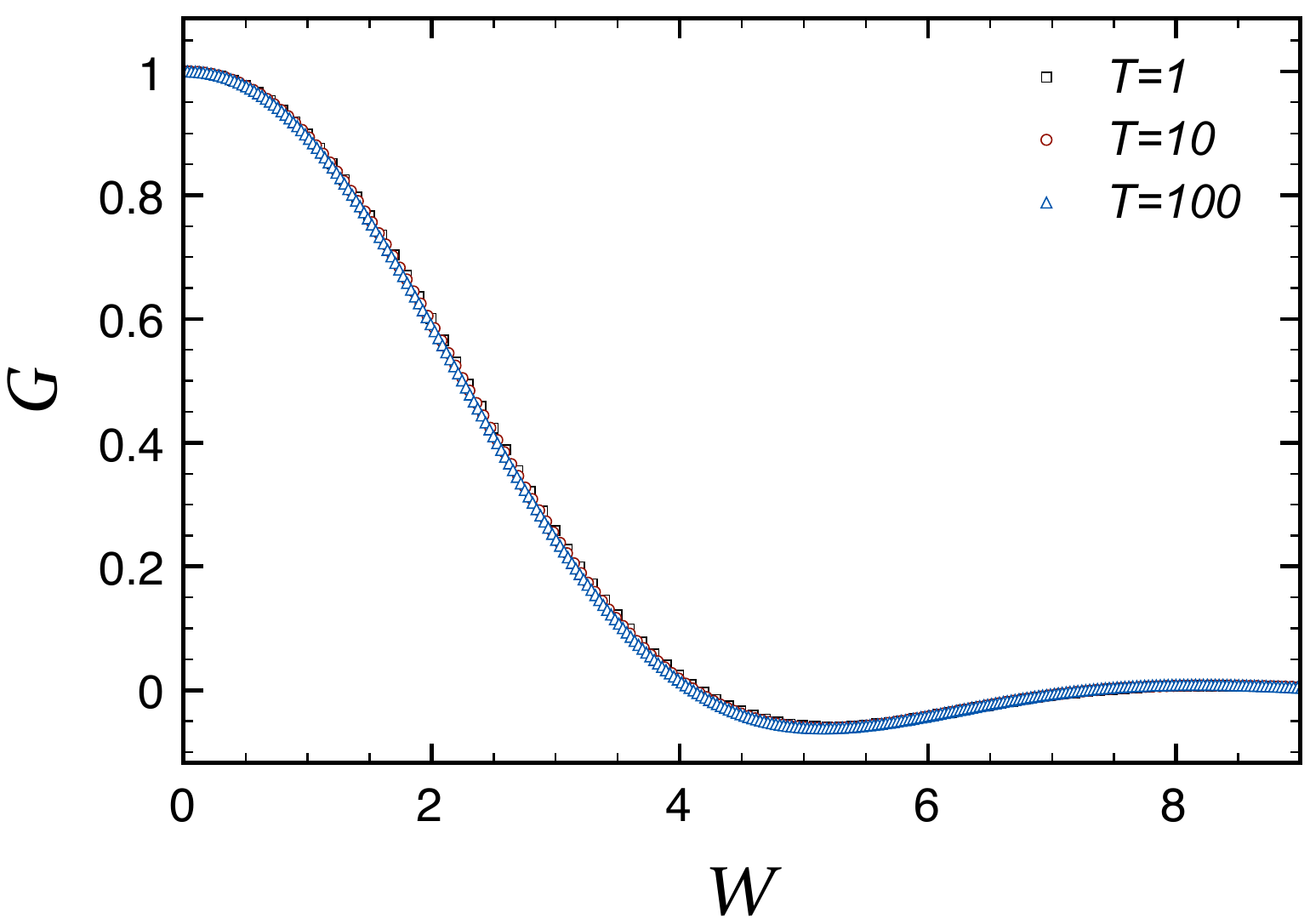}
\caption{\textit{Long-term numerical solution of Eq.~(\ref{adctfe}), after several dimensionless times, in the self-similar variables of Eq.~(\ref{ssc}). The initial condition is a spherical cap of dimensionless height $D_\textrm{c}=0.1$ and radius of curvature $R_\textrm{c}=2$, on a film with height $E=1$.}}
\label{dropSS}
\end{center}
\end{figure}

\begin{figure}
\begin{center}
\includegraphics[width=8.9cm]{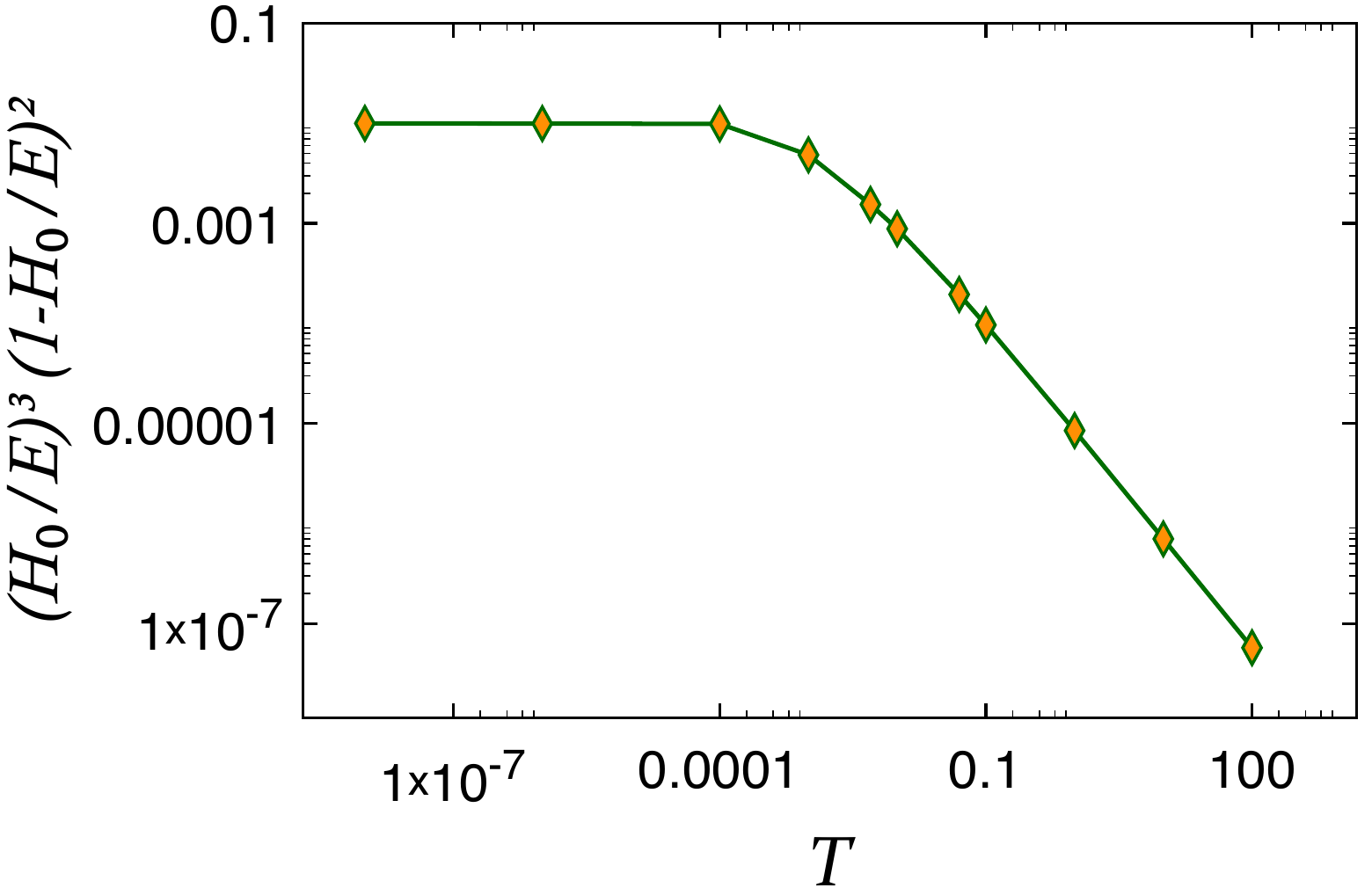}
\caption{\textit{Crossover between the two regimes of a leveling viscous droplet onto an identical film. Data was computed according to the numerical solution of Eq.~(\ref{adctfe}) and the function of $H_0$ introduced in Eq.~(\ref{model}).}}
\label{dropSScv}
\end{center}
\end{figure}

Guided by the symmetry of Eq.~(\ref{adctfe}) and Eq.~(\ref{cictfe}), we look for self-similar solutions of the second kind \cite{Barenblatt1996} defined by:
\begin{subequations}
\label{ssc}
\begin{align}
G(W)&=\frac{H(R,T)-E}{H(0,T)-E}\\
W&=\frac{R}{[H(0,T)^{3}\ T]^{1/4}}\ .
\end{align}
\end{subequations}
We thus plot the long-term evolution of the droplet of Fig.~\ref{dropfall} with these new variables. The result is shown in Fig.~\ref{dropSS}.
All the curves collapse onto a single curve, demonstrating the self-similarity of the long-term evolution. In order to characterize the latter, we introduce the additional dimensionless volume $\mathcal{V}$ with respect to the volume of the underlying film. It is constant by incompressibility of the liquid, and is defined by:
\begin{equation}
\label{vol}
\mathcal{V}=2\pi \int_0^{\infty}dR\ R\ [H(R,T)-E]\ ,
\end{equation}
which can be evaluated at $T=0$ since we know the initial profile. Using Eq.~(\ref{ssc}) and Eq.~(\ref{vol}), we derive the temporal evolution of the droplet height $H_0=H(0,T)$:
\begin{equation}
\label{model}
\left(\frac{H_0}{E}\right)^3\left(1-\frac{H_0}{E}\right)^2=\frac{\mathcal{V}^2}{\alpha^2T}\ ,
\end{equation}
where: 
\begin{equation}
\alpha=2\pi\int_0^{\infty}dW\ W\ G(W)\ .
\label{alph}
\end{equation}
The last quantity is a geometrical factor depending only on the dimensionless height $D_\textrm{c}$ and radius of curvature $R_\textrm{c}$ of the initial spherical cap. However, Eq.~(\ref{model}) and thus self-similarity of Eq.~(\ref{ssc}) appear to be only true at large times, as in \cite{Christov2012}. During the short-term evolution, the drop height does not change significantly since the Laplace pressure gradients are much greater near the initial contact line. Therefore, we expect a crossover between a transient regime of constant height during which the contact line at the periphery of the droplet becomes smooth - and the influence of the initial condition is vanishing - and the intermediate asymptotics of Eq.~(\ref{ssc}). The full evolution of $H_0$ is shown in Fig.~\ref{dropSScv}. It confirms the existence of two regimes. After $T\sim0.1$, the evolution is self-similar. Moreover, the long-term slope of the curve in Fig.~\ref{dropSScv} is self-consistent with the $\alpha$ and $\mathcal{V}$ values of Eq.~(\ref{model}), for each configuration that we tested numerically. 

\subsection{Comparison with experiments}
\begin{figure}
\begin{center}
\includegraphics[width=8.5cm]{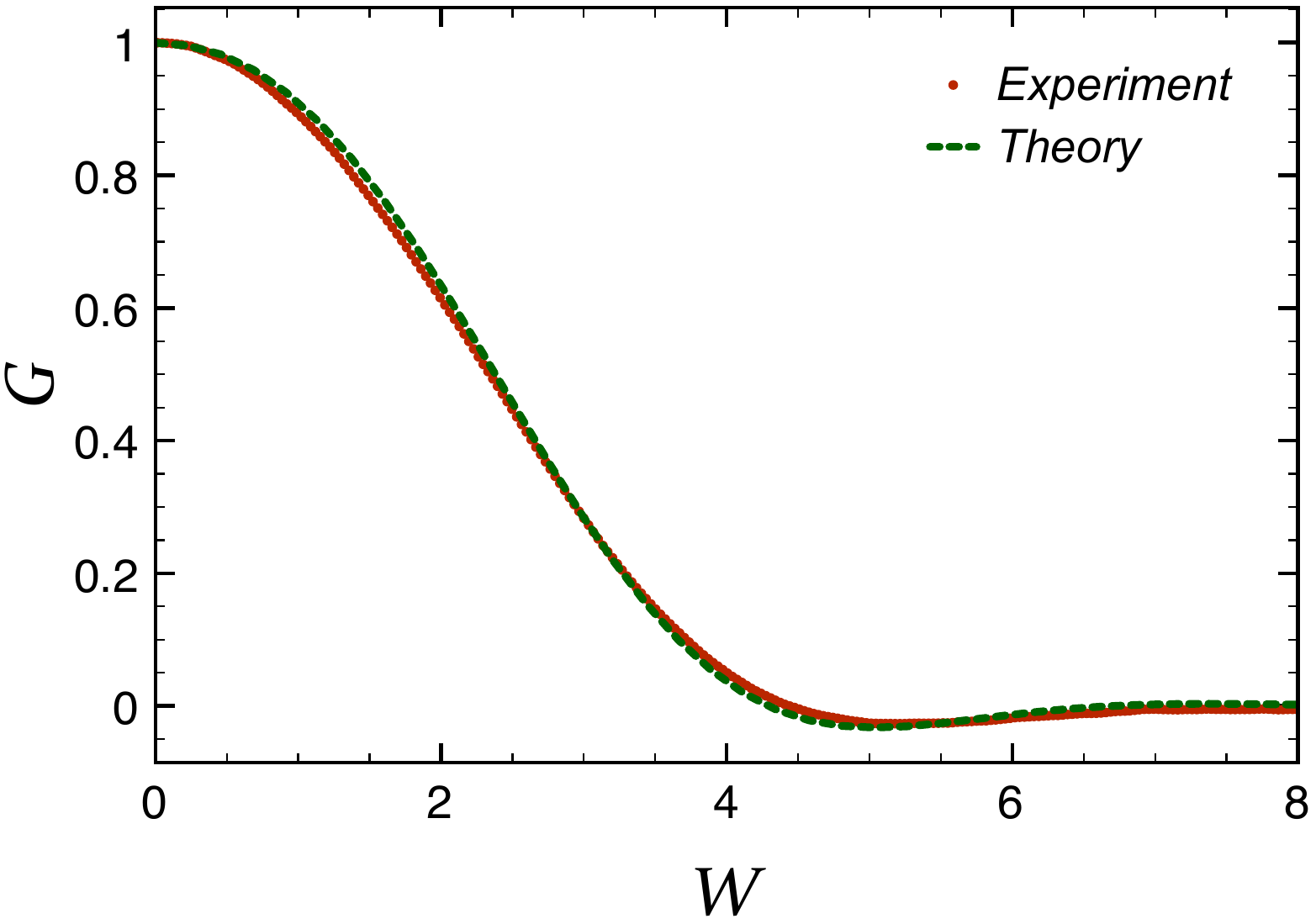}
\caption{\textit{Fit of the numerical solution of Eq.~(\ref{adctfe}) to the experimental profile (atomic force microscopy), in self-similar variables according to Eq.~(\ref{ssc}). The system is a thin polystyrene droplet with $M_\textrm{w}=118$~kg.mol$^{-1}$ leveling on an identical film, at $140\ ^{\circ}\textrm{C}$~\cite{Cormier2012}. The dimensionless parameters are: $T=105$, $D_\textrm{c}=0.98$, $R_\textrm{c}=67.5$ and $E=1$.}}
\label{resdrop}
\end{center}
\end{figure}

In order to validate the previous computational results, we compare the numerical solution to experimental profiles. The system is a thin polystyrene droplet with $M_\textrm{w}=118$~kg.mol$^{-1}$ leveling on an identical thin film, at $140\ ^{\circ}\textrm{C}$. Details can be found in~\cite{Cormier2012}, where we use the present numerical solution to establish the connection between Tanner's regime of a droplet spreading onto an infinitesimal precursor film \cite{Tanner1979,deGennes1985,deGennes2003,Bonn2009} and the opposite regime of a small droplet leveling into a thicker film. A typical result is shown in Fig.~\ref{resdrop}. Note that the self-similarity of the experimental profile has been verified. There is no free parameter in this comparison since the capillary velocity has been extracted from the leveling profile of a stepped film (see Section~\ref{sectfe}) with the same liquid and temperature \cite{McGraw2012}. 

\section*{Conclusion}
We presented the numerical implementation of capillary-driven thin film equations that describe the evolution of viscous films in two-dimensional configurations. In both cases, we presented the algorithm, the solution and its intermediate asymptotics that we compared to experimental data. 

The first application was the leveling of a step at the free surface of a thin polymer film. In this case, we demonstrated the self-similarity of the long-term evolution and characterized the robustness of this intermediate asymptotics of the first kind. Then, we compared the numerical solution to experimental profiles. The excellent agreement validates this technique as a useful viscometer for polymer melts. 

The second application that we discussed was the leveling of a liquid polymer droplet onto an identical film. In that case, we found another type of self-similar evolution. This intermediate asymptotics of the second kind, as well as the convergence to this regime, were characterized as in the case of the leveling of a stepped film. The evolution of a given droplet appeared to crossover from a transient regime of constant height where the initial contact line becomes smooth to the long-term self-similar regime. The agreement with experiments is also excellent. Whereas Tanner's regime of a droplet on an infinitesimal precursor film is well understood~\cite{Tanner1979,deGennes1985,deGennes2003,Bonn2009}, the numerical approach presented here offers a way to explore further the regime where the thickness of the underlying film is no longer negligible with respect to the height of the droplet~\cite{Cormier2012}. 

In near future, the computations may be extended to other geometries, fluids, and length scales. Disjoining pressure at the smallest scales~\cite{Seemann2001} or gravity at much larger scales~\cite{Huppert1982} may also be included. It is our hope that the simple approach presented here may be more generally used to probe rheology and boundary conditions on the nanoscale, and facilitate an understanding of hydrodynamics on length scales where continuum treatments break down. 

\section*{Acknowledgments}
The authors would like to thank Howard Stone and Jens Eggers for useful references and discussions. They thank as well the \'Ecole Normale Sup\'{e}rieure of Paris, the Natural Sciences and Engineering Research Council of Canada, the German Research Foundation (DFG) under grant BA 3406/2, the Chaire Total-ESPCI and the Saint Gobain Fellowship for financial support.

\end{document}